# Physics-informed neural network method for modelling beam-wall interactions

Kazuhiro Fujita[1]

[1] *Department of Information Systems, Saitama Institute of Technology, 1690 Fusaiji, Fukaya City 369-0293, Japan*

Email: kfujita@sit.ac.jp

A mesh-free approach for modelling beam-wall interactions in particle accelerators is proposed. The key idea of our method is to use a deep neural network as a surrogate for the solution to a set of partial differential equations involving the particle beam, and the surface impedance concept. The proposed approach is applied to the coupling impedance of an accelerator vacuum chamber with thin conductive coating, and also verified in comparison with the existing analytical formula.

*Introduction:* A relativistic beam of charged particles traversing in a particle accelerator can interact with vacuum chambers with resistive walls [1]. To estimate this interaction effect, which can limit the performance of a particle accelerator, the coupling impedance [2] is used. A typical way to obtain the coupling impedance is to use purely numerical methods. However, the volume mesh-type methods may suffer from numerical errors [3] in calculating the space-charge impedance. Calculating the *resistive-wall* impedance requires properly modelling the skin effect [1] in volume meshes. Therefore, it is still challenging to model the wideband impedance of vacuum chambers in accelerator designs.

We propose a novel mesh-free approach for modelling the resistive-wall impedance of accelerator vacuum chambers. The key idea of our method is to use a deep neural network (DNN) as an approximate solution to a set of partial differential equations (PDEs) involving the particle beam, and the surface impedance concept. The mesh-free feature originates from the use of the DNN, which has the universal function approximation capability. Our solution is based on the deep learning, inspired by the *physics-informed neural network* (PINN) [4].

This letter first introduces the PINN into the resistive-wall impedance modeling. The main focus is to extend our recent study [5] on the perfectly electric conducting (PEC) wall to the resistive wall. This approach is not yet addressed in other publications [6,7].

*Method:* The coupling impedance $Z_{\|}$ is defined in frequency domain as [2]

$$Z_{\|} = -\frac{E_z}{I} \qquad (1)$$

where $I=Qv$ is the total beam current, $Q$ is the total charge, $v=v\mathbf{e}_z$ is the beam velocity, $\mathbf{e}_z$ is the unit vector in the direction of beam motion ($z$-direction). To compute the impedance (1), we need to know the longitudinal component of the electric field $E_z$ for one particular harmonic component with an angular frequency $\omega=2\pi f$ (or wave number $k=\omega/v$). We assume that the beam has a rigid charge density distribution normalized by $Q$, and moves along the axis of an infinitely long vacuum chamber.

For the above beam-wall system, using the *special scaling scheme* [5] for deep learning, we can derive the following PDE

$$\left(\frac{\partial^2}{\partial X^2} + \frac{\partial^2}{\partial Y^2}\right)e_z - \frac{s_0^2 k^2}{\gamma^2}e_z + jB\rho_n = 0 \qquad (2)$$

where $(X,Y)=(x/s_0, y/s_0)$ are Cartesian coordinates scaled with a typical chamber length $s_0$ (e.g., radius, height and width), $e_z=E_z/E_0$ is the electric field scaled with

$$E_0 = \frac{s_0^2 kq_n}{B\varepsilon_0\gamma^2}, q_n = \frac{Q}{2\pi\sigma_x\sigma_y} \qquad (3)$$

and $\varepsilon_0$ is the permittivity of vacuum, $\gamma=(1-\beta^2)^{-1/2}$ is the relativistic factor, $\beta=v/c$, $B$ is an empirical parameter, $\rho_n$ is the normalized bi-Gaussian charge density

$$\rho_n = e^{-\frac{(x-x_c)^2}{2\sigma_x^2} - \frac{(y-y_c)^2}{2\sigma_y^2}} \qquad (4)$$

where $(\sigma_x,\sigma_y)$ is the half value of the Gaussian distribution in the $x$- and $y$-direction and $(x_c,y_c)$ is the center position in the transverse plane.

Here, we replace the original problem with resistive walls by an equivalent problem with the Leontovich boundary condition or *surface impedance boundary condition* (SIBC) [1]

$$E_z = -Z_s H_t \qquad (5)$$

where $H_t$ is the tangential component of the magnetic field on the surface of the chamber cross section, and $Z_s$ is the surface impedance function. Using (3), we can scale (5) as

$$e_z + Z_s H_t/E_0 = 0 \qquad (6)$$

Note that (6) is enforced only on the innermost wall of the chamber. All the domain outside the innermost wall is assumed to be filled by PEC. This can be also regarded as the assumption of infinitely thick PEC wall. Therefore, the field is zero outside the innermost chamber wall. In accelerator physics, this surface impedance concept can be used to reasonably model multiscale features of multiple surface perturbations such as the skin effect and the thin layer in the numerical method [8]. When $Z_s=0$, (6) can be reduced to just the PEC-BC $e_z=0$.

A schematic of the method is illustrated in Fig.1. The PDE (2) and the scaled SIBC (6) are involved into the loss function of a NN using automatic differentiation. This works well especially for smooth transverse charge density as in (4). Note that a neural network is used as a solution surrogate. Unlike the previous study [5], although the space-charge field has only a purely imaginary part, the resistive-wall wake field has both real and imaginary parts. Therefore, the constructed NN also has two outputs $(e_z^r, e_z^i)$ corresponding to the real ($r$) and imaginary ($i$) parts of $E_z$. Our algorithm is summarized in the following list.

1. Set up a computational domain surrounded by the innermost wall of a chamber cross section, the scaled SIBC (6), the physical constant $\varepsilon_0$, the beam parameters ($Q$, $v$, $\beta$, $\gamma$), a source domain related to $\rho_n$ and $q_n$, the wave number $k$, and the scaling parameters $s_0$ and $E_0$. Assume that the beam traverses inside the chamber and the field is zero outside the computational domain.

2. Generate randomly sampled points (or regular or irregular grid points) within the computational domain surrounded by the innermost wall. Note that no sampling point is generated outside the domain. The generated sampling points are used to train a NN as input.
3. Construct a NN with two outputs $\hat{e}_z^r(x,y;\theta), \hat{e}_z^i(x,y;\theta)$ as a surrogate of the scaled PDE solution $e_z(x,y) = e_z^r(x,y) + je_z^i(x,y)$, where $\theta$ is a vector containing all weights $w$ and bias $b$ in the neural network to be trained, $\sigma$ denotes an activation function.
4. Define the loss function $L$ defined by (2) and (6)
5. Train the constructed NN to find the best parameters $\theta$ by minimizing $L$ via the L-BFGS algorithm [9] as a gradient-based optimizer, until $L$ is smaller than a threshold $\epsilon$.
6. Obtain the coupling impedance (1) and original outputs (unscaled) from $\hat{E}_z(x,y;\theta) = E_0 \hat{e}_z(x/s_0, y/s_0;\theta)$ using the trained NN.

In this method, the loss function $L$ is defined by

$$L = w_{PDE} l_{PDE} + w_{SIBC} l_{SIBC} \quad (7)$$

$$l_{PDE} = \frac{1}{N_{PDE}} \sum_{p=1}^{N_{PDE}} \left[ |f_r(x_p, y_p; \theta)|^2 + |f_i(x_p, y_p; \theta)|^2 \right] \quad (8)$$

$$l_{SIBC} = \frac{1}{N_{SIBC}} \sum_{p=1}^{N_{SIBC}} \left[ |g_r(x_p, y_p; \theta)|^2 + |g_i(x_p, y_p; \theta)|^2 \right] \quad (9)$$

$$f_r = \left(\frac{\partial^2}{\partial X^2} + \frac{\partial^2}{\partial Y^2}\right)\hat{e}_z^r - \frac{s_0^2 k^2}{\gamma^2}\hat{e}_z^r \quad (10)$$

$$f_i = \left(\frac{\partial^2}{\partial X^2} + \frac{\partial^2}{\partial Y^2}\right)\hat{e}_z^i - \frac{s_0^2 k^2}{\gamma^2}\hat{e}_z^i + B\rho_n \quad (11)$$

$$g_r = \hat{e}_z^r + \text{Re}(Z_s H_t/E_0) \quad (12)$$

$$g_i = \hat{e}_z^i + \text{Im}(Z_s H_t/E_0) \quad (13)$$

where $p$ denotes the sampling point. $N_{PDE}$ and $N_{SIBC}$ are the numbers of sampling points in the computational domain and on the boundary surface, respectively. $w_{PDE}$ and $w_{SIBC}$ are the weights of the loss function. $l_{PDE}$ is the loss function related to the scaled PDE (2), and its minimization ($l_{PDE} \to 0$) enforces (2) at a set of finite sampling points in the computational domain. $l_{SIBC}$ is the loss function related to the SIBC, and its minimization ($l_{SIBC} \to 0$) enforces (6) at a set of finite sampling points on the boundary surface.

Throughout this study, we adapted a fully connected neural network and the tanh activation function. We used three hidden layers and 20 neurons per layer. We chose $B$=100, ($w_{PDE}$, $w_{SIBC}$)=(1,100) and ($N_{PDE}$, $N_{SIBC}$)=(2000,200). $N_{PDE}$ random points are generated inside a chamber and $N_{SIBC}$ grid points are generated on the chamber wall.

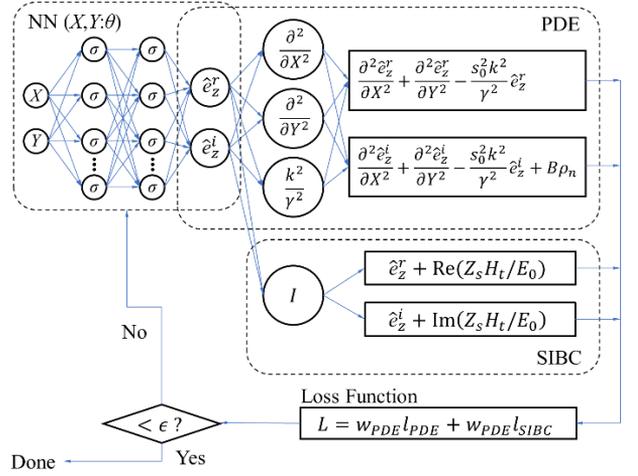

**Fig 1** Physics-informed neural network.

The prediction accuracy of this method depends on the NN architecture and hyperparameters in deep learning, as confirmed in [5]. Its general trend on accuracy [4] is that a good prediction accuracy can be achieved as a sufficiently expressive NN architecture and sufficient numbers of sampling points are given.

*Results and discussion:* To show the feasibility of the proposed method, we apply it to the analysis of a round vacuum chamber with inner radius $R$=25mm and the first layer with a small conductivity $\sigma$=400S/m and a thickness of 5mm followed by a PEC, as shown Fig.2(a). The surface impedance function on the innermost chamber wall can be given by [10]

$$Z_s(\omega) = j\eta_{ct} \tan k_{ct} d, k_{ct} = \omega\sqrt{\hat{\mu}_{ct}\hat{\varepsilon}_{ct}}, \eta_{ct} = \sqrt{\frac{\hat{\mu}_{ct}}{\hat{\varepsilon}_{ct}}} \quad (14)$$

$$\hat{\mu}_{ct} = \mu_0, \hat{\varepsilon}_{ct} = \varepsilon_0 - j\frac{\sigma}{\omega} \quad (15)$$

where $\mu_0$ is the permeability of vacuum. Note that the real part is different from the imaginary one in low frequencies. In this chamber, a round Gaussian beam with $Q$=1pC, $\gamma$=27.7 and $\sigma_x$=$\sigma_y$=$\sigma_r$=2.5mm traverses on its center.

Since the closed-form exact solution in this beam-wall system is not available, we verify our simulation with the PINN in comparison with the analytical impedance of a round beam with uniform transverse charge density

$$\rho_\perp = \begin{cases} \frac{Q}{\pi r_b^2} & r \leq r_b \\ 0 & r > r_b \end{cases} \quad (16)$$

$$r = \sqrt{(x-x_c)^2 + (y-y_c)^2}$$

in the circular chamber [11] involving the same surface impedance as (14). Here we choose the uniform beam radius as $r_b$=1.747$\sigma_r$. This choice allows us to approximate the field of the round Gaussian beam by that of the round uniform beam, up to $k\sigma_r/\gamma$≈0.5 as used in [5]. This condition reads $f$<0.66THz in our case.

Here, we assume that the magnetic field on the resistive wall is the same as that on the PEC wall at the same radius $R$=25mm. This hypothesis is often used for the impedance theory [1,10,11] in accelerator physics. We use the same magnetic field in calculating (6).

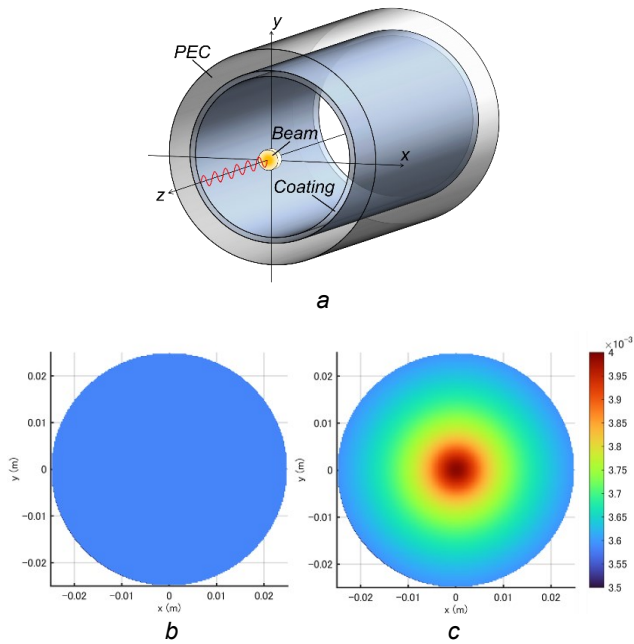

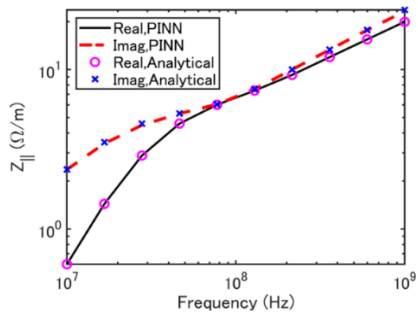

**Fig 2** Electric field distribution of a relativistic particle beam in infinitely long round vacuum chamber with thin conductive coating. (a) chamber geometry, (b) real part, (c) imaginary part.

**Fig 3** Comparison of coupling impedances of two different beams in the same round chamber with thin coating.

Fig.2(b),(c) shows the PINN-simulated result of the field ($E_z$) in the chamber at $f=3.6\times10^8$Hz. It is seen that the imaginary part of the field has a nonuniform distribution, and it becomes stronger near the chamber center. This main contribution originates from the space charge field, which has the purely imaginary part related to the charge density in the right-hand side of Eq. (2). By contrast, the real part of the field is nonzero, and it has a uniform distribution. It seems that the real part does not depend on $r$, and the field value at the axis is same as the one on the inner chamber wall. The similar field behavior is well explained in [1]. This indicates that the proposed method can model *both* the resistive-wall wake field and space-charge field.

Fig.3 shows the simulated impedances of the round Gaussian beam and the analytical impedance of the round uniform beam with $r_b=1.747\sigma_r$ in the same round chamber. Both for the real and imaginary parts, the simulated and analytical impedances is in excellent agreement. We can also see that the frequency dependency of the real part is different from that of imaginary part.

The results demonstrate that the constructed PINNs can properly model the beam-wall interactions addressed here and accurately simulate the fields and coupling impedance.

*Conclusion:* The PINN method with the surface impedance concept has been proposed for the coupling impedance modelling of accelerator vacuum chamber with resistive walls. To verify the method, a round chamber with thin conductive coating is analyzed. The computed coupling impedance agrees well with the approximated analytical impedance. Although only the round geometry with a single layer has been analyzed, the presented method can be extended to multilayer vacuum chambers with other geometries.

**References**
1. Stupakov, G., Penn, G.: Classical mechanics and electromagnetism in accelerator physics, Springer International Publishing AG, Cham, Switzerland, Chap.12, (2018)
2. Zotter, B., Kheifets, S.A.: Impedance and wakes in high-energy accelerators, World Scientific Publishing, Singapore, (1998)
3. Niedermayer, U., Boine-Frankenheim, O., De Gersem, H.: Space charge and resistive wall impedance computation in the frequency domain using the finite element method, Phys. Rev. ST Accel. Beam, **18**, 032001, (2015)
4. Raissi, M., Perdikaris, P., Karniadakis, G.E.: Physics-informed neural networks: A deep learning framework for solving forward and inverse problems involving nonlinear partial differential equations, J. Comput. Phys. **378**, 686–707, (2019)
5. Fujita, K.: Physics-informed neural network method for space charge effect in particle accelerators, IEEE Access, **9**, pp.164017-164025, (2021)
6. Edelen, A.L., Biedron, S.G., Chase, B.E., Edstrom Jr., D, Milton, S.V., Stabile, P.: Neural Networks for modeling and control of particle accelerators, IEEE Trans. Nucl. Sci., **63**(2), 107-119, (2016)
7. Ivanov, A, Agapov, I.: Physics-based deep neural networks for beam dynamics in charged particle accelerators, Phys. Rev. Accel. Beams, **23**, 074601, (2020)
8. Fujita, K.: Modeling of beam-wall interaction in a finite-length metallic pipe with multiple surface perturbations, IEEE J. Multiscale. Multiphys. Comput. Techn., **2**, 237-242, (2017)
9. Liu, D.C., Nocedal, J.: On the limited memory BFGS method for large scale optimization, Mathematical programing, **45**, 503-528, (1989)
10. Migliorati, M., Palumbo, L., Zannini, C., Biancacci, N., Vaccaro, V.G.: Resistive wall impedance in elliptical multilayer vacuum chambers, Phys. Rev. Accel. Beams, **22**, 121001, (2019)
11. Al-khateeb, A.M., et al: Analytical calculation of the longitudinal space charge and resistive wall impedance in a smooth cylindrical pipe, Phys. Rev. E, **63**, 026503, (2001)